\documentstyle[12pt]{article}                  
\makeatletter
\def\nothing#1{}
\newdimen\earraycolsep
\renewcommand{\thetable}{\arabic{table}}
\renewcommand{\thefigure}{\arabic{figure}}
\renewcommand{\title}[1]{%
  \vspace*{120\p@}%
  {\parindent \z@ \raggedright \reset@font
    \bfseries #1\par
    \nobreak
    \vskip 36\p@
  }}
\def\author#1{{\pretolerance=10000 \raggedright \advance \leftskip by 1in
 \noindent #1 \vskip 1pc}}
\def\affiliation#1{{\advance\leftskip by 1in \noindent #1 \vskip -1pc}}
\def\refnote#1{{$^{\hbox{\scriptsize #1}}$}}
\def\affnote#1{\llap{$^{\hbox{\scriptsize #1}}$}}

\renewcommand\section{\@startsection{section}{1}{\z@}{2pc \@plus 
      1ex minus .2ex}{1pc \@plus .2ex}{\reset@font
      \normalsize\bfseries\noindent
      {\addtocounter{section}{1}}\arabic{section}\ 
      {\setcounter{subsection}{0}
      \setcounter{subsubsection}{0}\setcounter{equation}{0}} }}
\renewcommand\subsection{\@startsection{subsection}{2}{\z@}{1pc \@plus 1ex
    minus.2ex}{1pc \@plus .2ex}
    {\reset@font\normalsize\bfseries
    \noindent{\addtocounter{subsection}{1}}%
    {\setcounter{subsubsection}{0}}\arabic{section}.\arabic{subsection}\ }}
\renewcommand\subsubsection{\@startsection{subsubsection}{3}{\parindent}
        {1pc \@plus 1ex minus.2ex}{-0.5em}{\reset@font\normalsize\bfseries%
        {\addtocounter{subsubsection}{1}} \hspace*{.6cm}
        \arabic{section}.\arabic{subsection}.\arabic{subsubsection}
        \hspace*{-7mm}}}

\def\AmS{{\protect\the\textfont2%
        A\kern-.1667em\lower.5ex\hbox{M}\kern-.125emS}}

\def\p@LaTeX{{\family{times}\series{m}\shape{n}\selectfont L\kern-.36em
\raise.3ex\hbox{\scriptsize A}\kern-.15em T\kern-.1667em\lower.7ex\hbox{E}
\kern-.125emX}}

\newlength{\colwidth}

\def\@oddhead{\hfil}
\def\@evenhead{\hfil}
\def\@oddfoot{{\bfseries\hfil\thepage}}
\def\@evenfoot{{\bfseries\thepage\hfil}}
\def\fnum@figure{\footnotesize\raggedright{\bfseries \figurename~\thefigure.}}
\def\fnum@table{\normalsize\raggedright{\bfseries \tablename~\thetable.}}
\long\def\@makecaption#1#2{\vskip 10\p@ {#1 #2\par}}
\long\def\@makefntext#1{\setbox0=\hbox{$\m@th^{\@thefnmark}$}
\noindent\hangindent=\wd0 \box0 #1}
\def\centerfig#1#2#3#4{\vspace*{#2}\relax
\centerline{\hbox to#1{\special{#4:#3.#4 x=#1, y=#2}\hfil}}}
\newbox\@atbox
\long\def\atable#1#2#3{\begin{table}[tbp]\centering\footnotesize
\setbox\@atbox\hbox{#2}
\parbox{\wd\@atbox}{\caption{#1}}\par\smallskip
#2
\par\smallskip\parbox{\wd\@atbox}{\raggedright #3}
\end{table}}
\def\@nbibitem#1{\noindent \hangindent=2pc \hangafter=1
\refstepcounter{enumi}\hbox to 2pc{\arabic{enumi}.\hfil}%
\immediate\write\@auxout{\string\bibcite{#1}{\arabic{enumi}}}}
\def\numbibliography{%
\section*{REFERENCES}%
\bgroup\footnotesize
\setcounter{enumi}{0}%
\def\newblock{\hskip .11em plus.33em minus.07em}%
\let\bibitem\@nbibitem}
\def\endnumbibliography{\par\egroup}
\makeatother
\begin{document}
\rightline{IHES/P/96/70~}
\rightline{hep-th/9610198}
\title{T-DUALITY AND THE MOMENT MAP\refnote{$\dagger$}}                       
           

\author{C. Klim\v c\'\i k,\refnote{1} P. \v Severa, \refnote{2}}  

\affiliation{\affnote{1}  C.N.R.S., I.H.E.S.,\\F-91440 Bures-sur-Yvette,\\
France\\ \affnote{2} Department of Theoretical Physics, Charles University,\\
V Hole\v sovi\v ck\'ach 2, CZ-180 00 Praha 8,\\Czech Republic}

\vskip3pc
\noindent $~^{\dagger}$
 {\small Based on talks given by the first author at the Argonne Summer Duality
Institute, 27 June -12 July 1996 and at the 
Carg\'ese School on Quantum Fields and
 Quantum Space-Time, 22 July - 3 August 1996.}

\begin{abstract}
Aspects of Poisson-Lie T-duality are reviewed in more algebraic
way than in our, rather geometric, previous papers. As a new result,
a moment map is constructed for the Poisson-Lie symmetry of the 
system consisting of open strings propagating in a Poisson-Lie group
manifold.
\end{abstract}

\section{INTRODUCTION}        

T-duality can be interpreted as the strong-weak coupling field
theoretical duality from the world-sheet point of view but also
as a discrete symmetry of the string theory from the space-time
one \cite{GPR}. In  the paper \cite{KS1}
we have argued that
the T-duality is in fact a manifestation of the well known 
duality in the category of Poisson-Lie groups at the classical
level and of Hopf algebras (quantum groups) at the quantum
level. We have shown how the structural features of the 
Abelian T-duality  can be encoded in the language of Poisson-Lie
groups, Drinfeld doubles, Lie bialgebras, Manin triples etc.
We have argued that this language, though it is not really necessary
tool for studying the Abelian case, can be directly used for the non-Abelian
generalizations of the T-duality where it becomes essential.

In a sense, we kept stressing the stringy applications of the formalism
in our previous works because we personally discovered the rich Poisson-Lie
world as a suitable tool for handling problems arising in  string
research. In the course of developping the Poisson-Lie T-duality program
we have often settled down our own terminology. This attitude looked quite
safe since, to our best knowledge, the Poisson-Lie groups have not been
applied previously in string theory. But after having got acquainted
better with the theory of integrable models, we found the
 Poisson-Lie world to be an extremely interesting structure per se, with
a well developped terminology. In particular, our
 notion of the `Poisson-Lie symmetry' of a $\sigma$-model
which we used  in \cite{KS1} is in clash with
the notion of the Poisson-Lie symmetry  of a general dynamical
system in the sense of \cite{D}.  Moreover, our notion
played an essential role in our formalism because the Poisson-Lie
symmetry was {\it the} property  required from a $\sigma$-model
in order the duality transformation on it could be performed.
We therefore felt a need
to clarify our string results in the form appropriate for Poisson-Lie
 experts.
 Having performed
 this exercise turned out to be fruitful not only from the terminological
point of view; in fact,  we have discovered the traditional
 Poisson-Lie symmetry
of our `Poisson-Lie' symmetric $\sigma$-models. Thus both notions are
intimately connected and we are devoting this article to a detailed
description of this connection. As often before, it turns out that
an elegant and well-understood structure in the Poisson-Lie world
finds a natural manifestation in the world of the Poisson-Lie
T-dualizable $\sigma$-models.

In the second (third)  section we describe our (traditional)
notion of the Poisson-Lie symmetry.
We characterize the interplay between the two  in section 4 where also
a moment map of the traditional Poisson-Lie symmetry is constructed
for the case of open strings in the backgrounds possessing the `new'
Poisson-Lie symmetry.

\section{``NEW'' POISSON-LIE SYMMETRY of $\sigma$-MODELS}

Consider a $2n$-dimensional group $D$ such that its Lie algebra ${\cal D}$
(viewed as a vector space) can be decomposed as the direct sum of
two subalgebras, ${\cal G}$ and $\tilde {\cal G}$, maximally isotropic
with respect to a non-degenerate invariant bilinear form on ${\cal D}$
\cite{D}. Such a group $D$ is referred to as the Drinfeld double.
If, moreover, each element of $D$ can be uniquely written  as the
product of two elements of the two groups $G$ and $\tilde G$ in both 
possible orders of $G$ and $\tilde G$ we shall refer to $D$ as to the 
`perfect' Drinfeld double. 
Of course ${\cal G}$ and $\tilde {\cal G}$ are 
the Lie algebras of $G$ and $\tilde G$, respectively and it is often said
that the groups $G$ and $\tilde G$ form  the double $D$.
Throughout this article, we shall work with the perfect doubles.

There exists a natural symplectic structure on the group manifold
$D$, first introduced by Semenov-Tian-Shansky in \cite{ST}.
It will play the crucial role in our presentation, therefore we devote
some place to the description of its properties. For doing that,
define $(\bigtriangledown_L f)_a,(\bigtriangledown_L f)^a,
(\bigtriangledown_R f)_a$
and $(\bigtriangledown_R f)^a$ as 
$$df=(\bigtriangledown_L f)_a(dll^{-1})^a+
(\bigtriangledown_L f)^a(dll^{-1})_a=$$
$$=(\bigtriangledown_R f)_a(l^{-1}dl)^a+(\bigtriangledown_R f)^a(l^{-1}dl)_a,
\eqno(1)$$
where $f$ is some function on the double and $l\in D$ parametrizes the group
manifold $D$. Clearly, the upper and lower indices for the forms $dll^{-1}$
(or $l^{-1}dl$) mean
$$dll^{-1}=(dll^{-1})_aT^a +(dll^{-1})^a\tilde T_a\eqno(2)$$
and $T^a$ and $\tilde T_a$ are the generators of ${\cal G}$ and 
$\tilde {\cal G}$, respectively, satisfying the duality property
$$\langle T^a,\tilde T_b\rangle=\delta^a_b.\eqno(3)$$
Needless to say, in all formulas $\langle .,.\rangle$ denotes 
the invariant bilinear form on the double.
Then the Semenov-Tian-Shansky Poisson bracket is given by
$$\{f,f'\}_D=(\bigtriangledown_L f)_a(\bigtriangledown_L f')^a-
(\bigtriangledown_R f)^a(\bigtriangledown_R f')_a\eqno(4)$$
for arbitrary functions $f,f'$ on the double.

Consider the functions $f,f'$ in (4) to be invariant with respect to the
right action of the group $\tilde G$ ($G$) on $D$. Then they can be
interpreted as functions on the group manifold $G$ ($\tilde G$)
and their Poisson bracket (4) defines a Poisson bracket on the group
manifold $G$ ($\tilde G$). This Poisson bracket can be written
as 
$$\{f,f'\}_{G}=\Pi_{ab}(g)(\bigtriangledown_L f)_a  
(\bigtriangledown_L f')_b,\eqno(5)
$$
where $\Pi(g)$ is certain
 antisymmetric tensor field  on $G$ whose explicit form can be easily
derived from (4).
It is given by
$$\Pi(g)=b(g)a(g)^{-1},\eqno(6)$$
where 
$$\langle g^{-1}T^i g, \tilde T_j\rangle\equiv  a(g)^i_j,
\qquad \langle g^{-1}\tilde T_i g, \tilde T_j\rangle\equiv b(g)_{ij}.
\eqno(7)$$ 
The
derivatives $(\bigtriangledown_L f)^a$ and $(\bigtriangledown_L f')^b$
in (5)   are defined with respect to the group manifold $G$.
Of course, the role of the groups $G$ and $\tilde G$ can be interchanged
and, up to a sign\footnote{The Poisson structure (4)
changes the sign upon exchanging $G$ and $\tilde G$ (cf. (28)).}, we 
obtain an exactly
 corresponding Poisson bracket 
 $\tilde \Pi^{ab}(\tilde g)$
on $\tilde G$. A pair of groups $G$ and $\tilde G$ equipped with the Poisson
brackets $\Pi(g)$ and $\Pi(\tilde g)$, respectively, is called a dual
pair of the Poisson-Lie groups.

In \cite{KS1}, we have constructed a dual pair of 
$\sigma$-models\footnote{These models
were shown to be dynamically equivalent (hence dual) in \cite{KS1}.}
on the group manifolds $G$ and $\tilde G$. Their Lagrangians were
respectively given as follows
$$L=E(g)^{ab}(\partial_+ gg^{-1})_a (\partial_- gg^{-1})_b;\eqno(8a)$$
$$\tilde L=\tilde E(\tilde g)_{ab}(\partial_+\tilde g\tilde g^{-1})^a 
(\partial_- \tilde g\tilde g^{-1})^b,\eqno(8b)$$
where
$$\partial_{\pm}\equiv\partial_{\tau}\pm\partial_{\sigma}\eqno(9)$$
and
$$E^{-1}(g)_{ab}=R^{-1}_{ab}+\Pi_{ab}(g),\quad \tilde E^{-1}(\tilde g)^{ab}=
R^{ab}
+\tilde\Pi^{ab}(\tilde g).\eqno(10)$$
Here $R$ is an arbitrary non-degenerate matrix. We have found in \cite{KS1},
that both models (8ab) are `Poisson-Lie symmetric' in the following sense:

\noindent {\it Definition:} A $\sigma$-model on a Poisson-Lie group
manifold $G$ is called Poisson-Lie symmetric if the Noether current
 one-forms $\tilde J(g)\in\tilde{\cal G}$  fulfil the zero-curvature condition
$$d\tilde J(g)-\tilde J(g)^2=0,\eqno(11)$$
for every solution $g$ of the $\sigma$-model field equations. Recall that
the Noether current one-forms $\tilde J(g)$ are defined by the variation
of the $\sigma$-model action with respect to the right action of $G$ on itself
$$\delta \int L=\int \langle \tilde J(g)\stackrel{\wedge}{,} d\epsilon\rangle
+\int \epsilon^a {\cal L}_{v_a}(L,)\eqno(12)$$
where $ \quad g+\delta g =g(1+\epsilon), \epsilon\in {\cal G}$ and
${\cal L}_{v_a}(L)$ are the Lie derivatives of the Lagrangian (see \cite{KS1}
for more details).

Note that  the models (8a) and (8b) are both Poisson-Lie symmetric;
 the role of the groups $G$ and $\tilde G$ in passing from (8a) to (8b) 
gets interchanged. 

\section{TRADITIONAL POISSON-LIE SYMMETRY}

Suppose we are given a manifold $P$ with the Poisson structure 
$\pi$ ($\pi\in \wedge^2 TP$)
and a right action $a: P\times G\to P$ of $G$ on $P$,
 generated by a section
$v$ of the bundle $TP\otimes {\cal G}^*=TP\otimes \tilde {\cal G}$. If we can
find a map $\tilde m: P\to  \tilde G$ such that
$$v= \pi(d\tilde m\tilde m^{-1})\eqno(13)$$
and $\tilde m$ is equivariant with respect to the action $a$ on $P$
and the dressing action of $G$ on $\tilde G$ then $\tilde m$ is called the 
moment map of the Poisson-Lie action $a$ of the Poisson-Lie group $G$ on the 
Poisson manifold $P$. Note that on this section we inheritate the 
terminology of the 
previous
one: $G$ and $\tilde G$ form the dual pair of the Poisson-Lie groups; moreover,
we understand that the dual spaces ${\cal G}^*$ and $\tilde {\cal G}^*$
are naturally identified with $\tilde {\cal G}$ and ${\cal G}$, respectively,
via the bilinear form $\langle .,.\rangle$ on the double. Recall also that
the 
definition of the dressing action of $G$ on $\tilde G$:
$$\tilde gh=g\tilde h,\quad g,h\in G, \quad \tilde g,\tilde h\in 
\tilde G.\eqno(14)$$
Here $h$ acts on $\tilde g$ and the result of the action is $\tilde h$. 
The fact that
the element of the double $\tilde gh$ can be uniquely represented as
the product $g \tilde h$ follows from our assumption that the Drinfeld double
 is perfect.

As an example of the moment map for a Poisson-Lie action, consider the 
Drinfeld double $D$ itself
as the manifold $P$; the Poisson
structure $\pi$ is given by the Semenov-Tian-Shansky bracket (4).
The action of the Poisson-Lie group $G$ on $D$ is given simply
by right multiplication of the elements of $D$ by the elements of its subgroup
$G$. The moment map for this action is given as
$$\tilde m(l)=\tilde h, \quad l\in D,\eqno(15)$$
where $\tilde h$ is given by the following decomposition of the 
arbitrary element $l\in D$:
$$  l=g\tilde h, \quad g\in G, \quad \tilde h\in \tilde G.\eqno(16)$$
The fact that $\tilde h$ is the moment map of this Poisson-Lie action can
be easily checked by direct computation (see also \cite{ST}).
Note, that the role of the groups $G$ and $\tilde G$ can be interchanged
and the right action of $\tilde G$ on $D$ is also the Poisson-Lie action whose
moment map can be constructed in the exactly corresponding way.

If the manifold $P$ is the symplectic manifold, which means that
the bivector field $\pi$ can be inverted to give the symplectic
form $\omega$ on $P$, then, as the corollary of (13), we have the
relation
$$i_v\omega=d\tilde m\tilde m^{-1}.\eqno(17)$$
Here $i_v$ means the contraction of the form by the vector field.
We say that a dynamical system, whose phase space is the symplectic
manifold $(P,\omega)$ is Poisson-Lie symmetric with respect to the group 
$G$ if 
$${\cal L}_v H=0,\eqno(18)$$
where $H$ is the Hamiltonian of the system 
and $v\in TP\otimes \tilde {\cal G}$ generates
the Poisson-Lie action of $G$ on $P$.

\section{A COMPARISON}
Consider an open string  in the Poisson-Lie group manifold
$G$ whose propagation is governed by (8a). The open string
boundary conditions at the world sheet boundaries $\sigma=0,\pi$
read
$$\partial_{\sigma}gg^{-1}\vert_{0,\pi}=0.\eqno(19)$$
They are usually referred to as the von Neumann conditions and
they insure that there is no momentum flow through the boundary of the
string. But here the `momentum' $\tilde m$ is a group valued quantity defined
for every extremal world sheet (=a point in the phase space of the 
system) by calculating the path-ordered integral of the Noether
current $\tilde J(g)$ over a path $\gamma$ starting at one edge of the open
 string
world sheet and ending at the other edge:
$$\tilde m=P\exp{\int_{\gamma}\tilde J(g)}.\eqno(20)$$
Note that the momentum $\tilde m$ is the conserved quantity; it does not
depend on the path $\gamma$ (in particular on the time in which 
$\gamma$ crosses the world sheet) by virtue of the equation (11) and 
the boundary conditions (19). Of course, we have a good reason to denote
the $\tilde G$-valued momentum as $\tilde m$; it is going to be precisely the
moment map (from the phase space of the system (8a) into the group $\tilde G$)
that generates the {\it traditional} Poisson-Lie symmetry of the model
(8a). In order to prove this, it is convenient to write the action of the
model
(8a) in the first-order (Hamiltonian) form \cite{KS2}
$$S=\int L_H,\eqno(21)$$
where
$$L_H=\langle \tilde\Lambda,g^{-1}\partial_{\tau}g\rangle
-{1\over 2}Ad_gG(\tilde \Lambda,\tilde \Lambda)$$
$$-{1\over 2}Ad_g G^{-1}
(g^{-1}\partial_{\sigma}g +Ad_g(B+\Pi(g))(.,\tilde\Lambda),
g^{-1}\partial_{\sigma} +Ad_g(B+\Pi(g))(.,\tilde\Lambda)).\eqno(22)$$
Here $\tilde\Lambda$ is the canonically conjugated
momentum that is naturally valued in ${\cal G}^*\equiv\tilde {\cal G}$. 
The bracket
$\langle .,.\rangle$ can therefore be understood in  two ways:
either as the standard pairing between algabra and coalgebra or
as the bilinear form in the Lie algebra ${\cal D}$ of the double $D$.
We have used a compact notation in (22) in order not to burden
the formula with too many indices: $G(.,.)$ and $B(.,.)$ are the symmetric
and the antisymmetric part, respectively,
of the bilinear form $R^{-1}(\tilde T_a,\tilde T_b)=(R^{-1})_{ab}$;
 $G^{-1}(.,.)$ is, in turn, the inverse bilinear form to $G(.,.)$ and, as such,
it is defined on the Lie algebra ${\cal G}$. $Ad_g$ means the adjoint action
of the group $G$ on the bilinear forms.

Our crucial trick is the following: we parametrize the canonically conjugated
momentum $\tilde \Lambda$ by a field $\tilde h$ (valued in $\tilde G$)
 such that
$$\tilde \Lambda=\partial_{\sigma}\tilde h\tilde h^{-1}.\eqno(23)$$
The ambiguity of this parametrization  is fixed by requiring that
$$\tilde h(\tau,\sigma=0)=\tilde e,\eqno(24)$$
where $\tilde e$ is the unit element of $\tilde G$. The first order
Lagrangian then can be  rewritten as
$$L_H=\langle \partial_{\sigma}\tilde h\tilde h^{-1},
g^{-1}\partial_{\tau}g\rangle -{1\over 2}\langle \partial_{\sigma}l~l^{-1},
A \partial_{\sigma}l~l^{-1}\rangle,\eqno(25)$$
where $l=g\tilde h$ and $A$ is a linear (idempotent) self-adjoint map
from the Lie algebra ${\cal D}$ of the double into itself. It has two 
eigenvalues $+1$ and $-1$, the corresponding eigensspaces ${\cal R}_+$ and 
${\cal R}_-$
have the same dimension dim$G$, they are perpendicular to each other in the 
sense of the invariant form on the double and they are given by the following 
recipe:
$${\cal R}_+={\rm Span}\{T^a +R^{ab}\tilde T_b\};\eqno(26a)$$
$${\cal R}_-={\rm Span}\{T^a -R^{ba}\tilde T_b\}.\eqno(26b)$$
The second term in the right hand side of (25) is (minus) Hamiltonian.
Obviously, the phase space of our model is described by the functions
$g$ and $\tilde h$, satisfying the boundary conditions (19) and (24).
Now we define an action of the group G on the phase space as follows
$$g\tilde hg_0=g'\tilde h',\eqno(27)$$
where $g_0\in G$ acts and the pair $g',\tilde h'$ is the result of the
action. We immediately notice that this action respects the boundary 
condditions to be fulfilled by $g$ and $\tilde h$. Moreover, the 
Hamiltonian is invariant for it can be written just as the function
of $\partial_{\sigma}ll^{-1}$ where $l=g\tilde h$. Does  this action of
$G$ on $P$ give rise to the Poisson-Lie symmetry of our open string 
 model  in the traditional sense of this notion? The answer is affirmative;
in order to prove this, we have to exploit again the properties of the
Semenov-Tian-Shansky symplectic structure on the double $D$.

We have proved in \cite{KS1}, that the Semenov-Tian-Shansky 
symplectic form $\omega$ on $D$ can be conveniently expressed as
$$\omega =\langle d\tilde h~\tilde h^{-1} \stackrel{\wedge} {,}g^{-1}
dg\rangle-\langle dh~h^{-1}\stackrel{\wedge}{,}
\tilde g^{-1}  d\tilde g\rangle,\eqno(28)$$
where we have used the  following two parametrization
of the group manifold $D$:
$$ l=g\tilde h, \qquad l\in D,\quad g\in G, \quad \tilde h\in \tilde G;
\eqno(29a) $$
$$ l=\tilde g h, \qquad l\in D, \quad \tilde g\in \tilde G,\quad h\in G.
\eqno(29b)$$
We already know from the previous section 
that if $v\in TP\otimes\tilde {\cal G}$
describes the infinitesimal right action of $G$  on $D$ then it holds
$$i_v\omega=d\tilde h \tilde h^{-1}, \qquad l=g\tilde h.\eqno(30)$$
Returning to our Hamiltonian first-order action (25), we observe
that the first term in it can be conveniently rewritten as
$$ \int_{\rho} d\tau d\sigma\langle 
\partial_{\sigma}\tilde h\tilde h^{-1},g^{-1}
\partial_{\tau}g\rangle=\int_{\rho} d\tau d\sigma\langle 
\partial_{\sigma} h~ h^{-1},
\tilde g^{-1}\partial_{\tau}\tilde g\rangle +\int_{\rho} ~^{*}\omega.
\eqno(31)$$
Here the $~^*\omega$ means the pull-back of the Semenov-Tian-Shansky form
on the world-sheet $\rho$ of the open string and we used both parametrizations
(29a) and (29b) of the double. Now it is obvious that under the
action $v$  of the group $G$  the variation of this quantity becomes
$$\delta\int d\tau d\sigma\langle 
\partial_{\sigma}\tilde h\tilde h^{-1},g^{-1}\partial_{\tau}g\rangle=
\delta\int ~^*\omega\equiv  
\int ~^*{\cal L}_v\omega.\eqno(32)$$
Because ${\cal L}_v=i_v d +d i_v$ and $\omega$ is closed, we have
$$\delta\int d\tau d\sigma\langle \partial_{\sigma}\tilde h
\tilde h^{-1},g^{-1}\partial_{\tau}g\rangle=\int d~^*i_v\omega=
\int_{\partial\rho}d\tilde h\tilde h^{-1}=$$
$$=\int_{\tau=\tau_f}d\sigma \partial_{\sigma}\tilde h\tilde h^{-1}-
\int_{\tau=\tau_i}d\sigma \partial_{\sigma}\tilde h\tilde h^{-1}
+\int_{\sigma=\pi}d\tau \partial_{\tau}\tilde h\tilde h^{-1}.\eqno(33)$$
Here $\tau_i(\tau_f)$ is some constant initial (final) time and the integral
along the edge $\sigma=0$ vanishes because of (24).
The formula (33) is almost what we want. The reason is that
the Hamiltonian first-order action is always of the form
$$S=\int_{\gamma}\alpha-Hdt,\eqno(34)$$
where $\alpha$ is the polarization form of the system (this means that
$d\alpha=\Omega$ and $\Omega$ is the symplectic form on the phase space
of the dynamical system) and $\gamma$ is a path in the phase space. 
In our case 
$$\int_{\gamma}\alpha=\int_{\rho} d\tau d\sigma\langle \partial_{\sigma}
\tilde h\tilde h^{-1},g^{-1}\partial_{\tau}g\rangle.\eqno(35)$$
If we want to show that the action (27) of $G$ on $P$
is the traditional Poisson-Lie
symmetry (the invariance of the Hamiltonian we have already shown),
  we have to prove that
$$i_v\Omega=d\tilde m\tilde m^{-1}\eqno(36)$$
for some $\tilde G$-valued function $\tilde m$ on the phase space $P$.
Note that 
$$i_v\Omega=i_vd\alpha={\cal L}_v\alpha -d(i_v\alpha) \eqno(37)$$
hence 
$$\int_{\gamma}i_v\Omega=\delta\int_{\rho}d\tau d\sigma\langle
\partial_{\sigma}\tilde h\tilde h^{-1},g^{-1}\partial_{\tau}g\rangle-
i_v\alpha\vert^{\tau_f}_{\tau_i}.\eqno(38)$$
Here $\tau_{i(f)}$ is the initial (final) time of the path $\gamma$.
By using the equation (33), we obtain
$$i_v\Omega =d\tilde h(\sigma=\pi)\tilde h^{-1}(\sigma=\pi),\eqno(39)$$
hence the moment map $\tilde m$ is just
$$\tilde m=\tilde h(\sigma=\pi).\eqno(40)$$
Finally, we observe
$$\tilde m=P\exp{\int_{\tau=const}\tilde \Lambda}=
P\exp{\int_{\tau=const}\tilde J(g)}.\eqno(41)$$
Thus, indeed, the $\tilde G$ valued momentum $\tilde m$, introduced
in (20), is the moment map of the traditional Poisson-Lie
symmetry of our open string $\sigma$-model (8a).


\section{ACKNOWLEDGEMENTS}    
C.K. thanks to the Argonne National Laboratory (in particular to C. Zachos)
for hospitality during the Institute  and to J. Schnittger for many 
discussions on related subjects.

\numbibliography 
\bibitem{GPR}{A.~Giveon, M.~Porrati and E.~Rabinovici,
Phys. Rep. 244 (1994) 77}
           
\bibitem{KS1}{C. Klim\v c\'\i k and P. \v Severa, Phys. Lett. B351 (1995)
455,
hep-th/9502122;  P. \v Severa,
{\it Minim\'alne  Plochy a Dualita},  Diploma thesis, Praha University, May
1995 (in Slovak);
C. Klim\v c\'\i k, Nucl. Phys. (Proc. Suppl.) 46 (1996) 116}

\bibitem{D}{F. Falceto and K. Gaw\c edzki, J. Geom. Phys. 11 (1993) 251;
A.Yu. Alekseev and A.Z. Malkin, Commun. Math. Phys. 162 (1994) 147}

\bibitem{ST}{M. A. Semenov-Tian-Shansky, Publ. RIMS, 
Kyoto Univ. 21 (1985) 1237}

\bibitem{KS2}{C. Klim\v c\'\i k and P. \v Severa, Phys. Lett. B372 (1996) 65}

\endnumbibliography

\end{document}